\documentclass[aps,pre,twocolumn]{revtex4}
\usepackage{amsmath,bm,epsfig}

\def\Fbox#1{\vskip1ex\hbox to 8.5cm{\hfil\fboxsep0.3cm\fbox{%
  \parbox{8.0cm}{#1}}\hfil}\vskip1ex\noindent}  

\let \= \equiv \let\*\cdot \let\~\widetilde \let\^\widehat \let\-\overline

\def\<{\left\langle}    \def\>{\right\rangle}
\def\({\left(}          \def\){\right)}
 \def \[ {\left [} \def \] {\right ]}

\begin{document}
\title{Non-Universality of the Specific Heat in Glass Forming Systems}
\author{ H.G.E. Hentschel$^*$, Valery Ilyin and Itamar Procaccia}
\affiliation{Department of Chemical Physics, The Weizmann
Institute of Science, Rehovot 76100, Israel \\
$^*$ Dept of Physics, Emory University, Atlanta Ga 30322. }
\date{\today}
\begin{abstract}
We present new simulation results for the specific heat
in a classical model of a binary mixture glass-former in two dimensions.  We show that in addition to the formerly observed specific heat peak there is a second peak at lower temperatures which was not observable in earlier simulations. This is a surprise, as most texts on the glass transition expect a single specific heat peak. We explain the physics of the two specific heat peaks by the micro-melting of two types of clusters. While this physics is easily accessible, the consequences are that one should not expect any universality in the temperature dependence of the specific heat in glass formers. 

\end{abstract}
\maketitle

The thermodynamic properties of glass-formers near the glass transition have been a subject of
intensive and far from settled research for more than half a century \cite{48Kau,68SS,76AS,85BN}. The temperature dependence of the entropy and how the entropy extrapolates to low temperatures gave rise
to the so-called Kauzmann paradox \cite{48Kau} that remains confusing to the present time. Important to the understanding of these issues is the specific heat, either at a constant volume or at a constant pressure, since its integral over a temperature path provides the entropy.
Experimental measurements of the specific heat in glass-forming systems are obtained from the linear response to either slow cooling (or heating) or to oscillatory perturbations with a given frequency about
a constant temperature. The latter method gives rise to a complex specific heat with the constraint that the zero frequency limit of the real part should be identified with thermodynamic measurements. Such measurements reveal anomalies in the temperature dependence of the specific heat, including the so called ``specific heat peak" in the vicinity of the glass transition. In fact, throughout the literature
on the glass transition one finds references to {\bf the} specific heat peak \cite{96EAN}. In this Letter we show that this concept must be discarded, since depending on the detailed physics of the system
there can be two or multiple specific heat peaks. We will present evidence for a model system with
two specific heat peaks, explain in detail the physical origin of the latter, and point out the important consequence that there is very little (or no) universality that can be expected in the
thermodynamic properties of different glass-formers. 

The model discussed here is the classical example \cite{89DAY,99PH} of a 
glass-forming binary mixture of $N$ particles in a 2-dimensional domain of 
area $V$, interacting via a soft $1/r^{12}$ repulsion with a `diameter' 
ratio of 1.4.  We refer the reader to the extensive work done on this system 
\cite{89DAY,99PH,07ABHIMPS,07HIMPS,07IMPS}. The sum-up of this work is that 
the model is a {\em bona fide} glass-forming liquid meeting all the criteria 
of a glass transition. 

In short, the system consists of an
equimolar mixture of two types of point-particles, ``large"  with interaction range
$\sigma_2=1.4$ and ``small" with interaction range $\sigma_1=1$, respectively, but 
with the same mass $m$. In general, the three pairwise
additive interactions are given by the purely repulsive soft-core potentials
\begin{equation}
\phi_{ab}(r) =\epsilon \left(\frac{\sigma_{ab}}{r}\right)^{n} \ , 
\quad a,b=1,2 \ ,
\label{epot}
\end{equation}
where $\sigma_{aa}=\sigma_a$ and $\sigma_{ab}= (\sigma_a+\sigma_b)/2$. The
cutoff radii of
the interaction are set at $4.5\sigma_{ab}$. The units of mass, length, time
and temperature are $m$, $\sigma_1$, $\tau=\sigma_1\sqrt{m/\epsilon}$ and
$\epsilon/k_B$, respectively, with $k_B$ being
Boltzmann's constant. In numerical calculations the stiffness parameter of the
potential (\ref{epot}) was chosen to be $n=12$. 

The isochoric specific heat is determined by the
fluctuations of the energy of the system at a given temperature:
\begin{equation}
\frac{C_{V}}{N}=\frac{d}{2}+\frac{\langle U^2\rangle-\langle U\rangle^2}{NT^2} \ . 
\label{cv1}
\end{equation}
The specific heat of our binary mixture model was measured at constant volume 
in \cite{99PH,02PH} and by us.  We have used the
last equation which allows one to estimate the specific heat in a single run 
of the canonical ensemble Monte Carlo simulations. At each
temperature the density was chosen in accordance with the simulation results
in an NPT ensemble as described in \cite{99PH} with the pressure value fixed
at $P=13.5$. As the initial configuration in the Monte Carlo process the last
configuration of the molecular dynamics run for this model at given
temperature after $1.3 \times10^8$ time steps was used. After short
equilibration the potential energy distribution functions were measured during
$2\times 10^6$ Monte Carlo sweeps. The acceptance rate was chosen to be $30\%$. 
Simulations were performed with $N=1024$ particles in a square cell with
periodic boundery conditions.
\begin{figure}[!h]
\centering
\hskip -2 cm
\includegraphics[width= 0.50\textwidth]{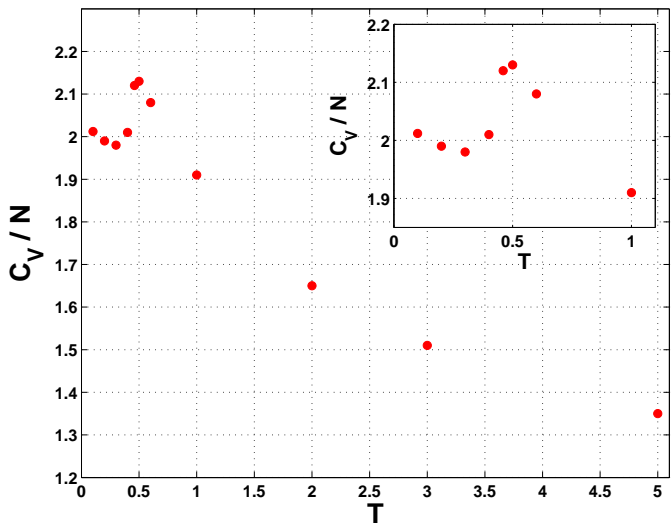}
\caption{Color online: in dots: the temperature dependence of the specific heat in the binary mixture model, computed at constant volume, such that the volume agrees with the pressure P=13.5 \cite{99PH} at each temperature.  The data indicate the existence of two specific heat peaks, one prominent at about $T=0.5$ and a smaller on at about $T=0.1$, and see the inset for finer detail. The data at lower temperature represent two years of computing time and are believed to be trustable.}
\label{Cfig3}
\end{figure}
 Our simulations appear to provide trustable values of $C_{V}$
down to lowest temperatures where the value of the specific heat coincides
with that of two-dimensional solid, i.e. $C_{V}=2$. What could not be seen in earlier simulations
is that there is a smaller second peak
of the specific heat at lower temperatures. To resolve it to the naked eye we present in Fig.
\ref{Cfig3} a blow-up of the region of lowest temperatures where the second peak
is more obvious. 

\begin{figure}
\centering
~\hskip -1.4 cm
\includegraphics[width=0.65\textwidth]{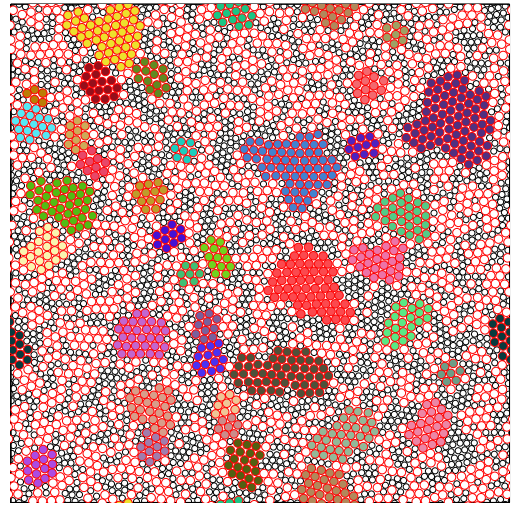}
\caption{(Color online). A snapshot of the system at $T=0.44$. In colours we highlight the clusters of large particles in  local hexagonal order. The colours have no meaning.}
\label{clusters}
\end{figure}

To understand the nature of the specific heat anomalies one must understand the physics that is behind the glassy behavior of this model in general and the existence of the two specific heat peaks in particular. When the temperature is lowered at a fixed pressure this system \cite{08LP} (as well as many other glass-formers \cite{06ST,09LPR,GS82,93Cha,98KT,00TKV}  tends to form micro-clusters of local order. In the present case 
large particles form hexagonal ordering first (starting at about $T=0.5$, and at lower temperatures (around $T=0.1$) also the small particles form hexagonal clusters. The clusters are not that huge,
with at most O(100) particles, (cf. Fig. \ref{clusters}), depending on the temperature and the aging time. But we have shown that the long time properties
of correlation functions are entirely carried by the micro-clusters \cite{08LP}. Below we will refer to the micro-clusters as curds and the liquid phase as whey.
We will argue that the specific heat responds to the micro-melting of the clusters - those of small particles at the lowest temperatures and those of the larger particles at higher temperatures. The large increase in
the number of degrees of freedom when a particle leaves a crystalline cluster and joins the liquid background is the basic reason for the increase in entropy that is seen as a specific heat peak.

We will calculate the specific heat at constant volume per particle from the exact expression that can
be derived \cite{09HIPS} for any system with inverse power law potential $r^{-n}$,
\begin{equation}
\frac{C_{V}}{N}=1+4\frac{K^{\infty}-K}{n^2\rho T}. \label{cv2}
\end{equation}
For our system $n=12$, $K$ is the bulk modulus and $K^\infty$ \cite{65ZM,LM06} is given by:
\begin{equation}
K^\infty=\rho T+\frac{n(n+2)}{4}\rho\frac{\langle U\rangle}{N}. \label{borns}
\end{equation}
The bulk modulus requires an
equation of state for its calculation. In the rest of this Letter we will therefore derive an approximate equation of state and compute the specific heat, exposing the origins of the two peaks. 

 To start we define $v_w^\ell$ , $v_w^s$, $v_c^\ell$ and $v_c^s$ respectively as the volume of large particle in the whey, small particle in the whey, large particle in the solid and small particle in the solid.
Similarly we denote by $ \epsilon_w^\ell$ , $\epsilon_w^s $, $ \epsilon_c^\ell$ and $\epsilon_c^s $ the energy of a large and small particle in the  in the whey and in the crystalline phase respectively.  Needless to say, all these quantities are temperature and pressure dependent; we will therefore explicitly use our low temperature knowledge concerning $v_c^\ell$ and $v_c^s$ in the crystalline phase, but treat the 
difference $v_w^\ell-v_c^\ell$ and $v_w^s-v_c^s$ as constants that we estimate below from our simulation knowledge. Similarly we estimate $\epsilon_c^\ell$ and $\epsilon_c^s$
from our knowledge of the hexagonal lattices at $T=0$. We assume that $\epsilon_w^\ell\approx \epsilon_c^\ell$ and similarly $\epsilon_w^s\approx \epsilon_c^s$ since our simulations indicate
a very small change in these parameters, see Table \ref{table}. It should be stressed that the enthalpy
change at these pressures are almost all due to the $PV$ term.
This will result in a semi-quantitative theory ascribing the important changes in specific heat to the changes in the fraction of particles in curds and whey. In other words the number of particles in the whey and the number of clusters are all explicit functions of temperature and pressure.
\begin{table}[h!]
\centering
\begin{tabular}{|c|c|c|c|c|c|c|c|}
\hline
$\epsilon_c^\ell$&$\epsilon_c^s$&$\epsilon_w^\ell$&$\epsilon_w^s$&$v_c^\ell$&$v_c^s$&
$v_w^\ell$&$v_w^s$\\
\hline
3.69&2.07&3.76&2.16&1.43&0.92&1.58&0.94\\
\hline
\end{tabular}
\caption{Parameters used in the calculation of the specific heat}
\label{table}
\end{table}

As the condensed phase consists of clusters of large and small particles, we use the notation $N_n^\ell$ for the number of clusters of $n$ large particles and $N_m^s$ for the clusters of $m$ small particles.   Here we only need the intensive variables $p_c^\ell=2\sum_nN_n^\ell/N$, $p_c^s=2\sum_mN_m^\ell/N$ $p_w^\ell=2N_w^\ell/N$ and $p_w^s=2N_w^s/N$ which stand for the fraction of large particles and small particles in the curds,  and large particles and 
small particles in the whey, such that $p_c^\ell+p_w^\ell=1$ and $p_c^s+p_w^s=1$. 
Using these variables we can write an expression for the volume per particle $v\equiv V/N$:
\begin{equation}
v=\frac{v_w^\ell+v_w^s}{2} +\frac{v_c^\ell-v_w^\ell}{2}p_c^\ell+\frac{v_c^s-v_w^s}{2}p_c^s \ . \label{vol}
\end{equation}

At this point we need to derive expressions for $p_c^\ell$ and $p_c^s$. To do so we need to 
remember that in the relevant range of temperatures the large particles in the whey can occupy
either hexagonal or heptagonal Voronoi cells, whereas small particles can occupy only pentagonal
or hexagonal cells \cite{07ABHIMPS,07HIMPS,08LP}. Accordingly there are  $g^\ell_w\approx (2^6 -1)/6+2^7/7$ ways to organize the neighbours of a large particle in the whey (neglecting the rare large particle in heptagonal neighbourhood), but only one way in the cluster. Similarly, there are  $g^s_w\approx (2^6-1) /6+2^5/5$ ways to organize a small particle in the whey. We note that this estimate assumes that the relative occurrence of the different Voronoi cells is temperature independent. While reasonable at higher temperatures \cite{08LP}, at lower temperature one should use the full statistical mechanics as presented in
\cite{07HIMPS} to get more accurate estimates of $g^\ell_w$ and $g^s_w$. This is not our aim here; we 
aim at a physical understanding of the specific heat peaks rather than an accurate theory. We thus
end up with the simple estimates
\begin{eqnarray}
p^\ell_c (P,T) 
&\approx& \frac{1}{1+g^\ell_w e^{[(\epsilon_c^\ell-\epsilon_c^\ell)+P (v_c^\ell-v_w^\ell)]/T}} \ , \label{pellc}\\
p^s_c(P,T)& \approx & \frac{1}{1+g^s_w e^{[(\epsilon_c^s-\epsilon_w^s)+P (v_c^s-v_w^s)]/T}} \ . \label{psc}
\end{eqnarray}

It is important to note that the combination of Eq. (\ref{vol}) together 
with Eqs. (\ref{pellc}) and (\ref{psc}) provides a mechanical equation of state.  We will now compute $C_v$ directly from Eq. (\ref{cv2}). The peaks in the specific heat are determined by the temperature dependence of $p^\ell_c (P,T) $
and $p^s_c (P,T)$ each which has a temperature and pressure derivatives that peaks at a different temperature, denoted as  $T^\ell(P)$ and $T^s(P)$. As said above
we take $\Delta v^\ell\equiv v_w^\ell-v_c^\ell$ and $\Delta v^s \equiv v_w^s-v_c^s$ as approximately constants (as a function of temperature and pressure). The constants are estimated from the condition that the second temperature derivative of 
 $p^\ell_c (P,T) $ and $p^s_c (P,T)$ should vanish. From this conditions we find
 \begin{equation}
 \Delta v^\ell\ \approx T^\ell(P^*) \ln g_w^\ell/P^*\ , \quad  \Delta v^s\ \approx T^s(P^*) \ln g_w^s/P^* \ ,
 \end{equation}
where $P^*$ is the pressure for which the peaks in the derivatives are observed (13.5 in our simulations). This is equivalent to a linear dependence of the specific heat peaks as a function
of pressure, $T^\ell(P)/T^\ell(P^*)=P/P^*$ and similarly for the small particles.
\begin{figure}
\centering
\includegraphics[width= 0.55\textwidth]{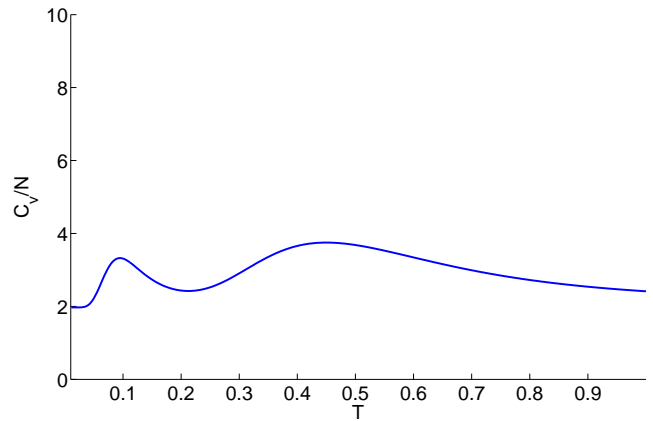}
\caption{Specific heat at constant volume as predicted by the simple theory which is based on
the mechanical equation of state supplied by Eqs. (\ref{vol}) and (\ref{pellc}) and (\ref{psc}). Note that the theory predicts the two peaks which are associated with the micro-melting or micro-freezing of the clusters of large and small particles respectively. The magnitude of the peaks is too high, reflecting terms missing in the simple approach, like the effect of anharmonicity at the lowest temperatures which are negative, tending to decrease the height of the low-temperature peak.}
\label{Cvtheory}
\end{figure}

In terms of these objects we can rewrite
\begin{eqnarray}
&&v =v_c(P,T)+  \Delta v^\ell(1-p_c^\ell)+ \Delta v^s (1-p_c^s) \ , \label{eqstate1}\\
&&\left(\frac{\partial v}{\partial P}\right)_T =\left(\frac{\partial v_c}{\partial P}\right)_T-  \Delta v^\ell \left(\frac{\partial p_c^\ell}{\partial P}\right)_T-\Delta v^s \left(\frac{\partial p_c^\ell}{\partial P}\right)_T \ .\label{eqstate2}
\end{eqnarray}
To compute the temperature dependence of $\left(\frac{\partial v}{\partial P}\right)_T$ we need first
to determine its $T\to 0$ limit, which is determined by the first term on the RHS of Eq. (\ref{eqstate1}) as the other terms on the RHS decay exponentially fast when $T\to 0$. Since we have already exact
results for the bulk modulus for the present model, we return
to Eqs. (\ref{cv2}) and (\ref{borns}).  We know on the one hand that $\lim_{T\to 0} C_v=2$ and that
$\langle U\rangle /N\approx 2.94$ over the whole interesting temprature range, cf. \cite{09HIPS}.
The compressibility $\kappa$  is related to the bulk modulus via
$\kappa=-\left(\frac{\partial v}{\partial P}\right)_T/v=1/K$ and therefore $\left(\frac{\partial v_c}{\partial P}\right)_T\approx -1/(123.5-35T)$ is easily estimated as $T\to 0$.  For simplicity we will use this approximation up to $T\approx 0.5$. 

Having all the ingredients we can compute $C_v/N$ . The parameters used were
estimated from the numerical simulation and are summarized in Table \ref{table}. Since the 
aim of this subsection is only semi-quantitative, we do not make any attempt of parameter fitting, and show the result of the calculation in Fig. \ref{Cvtheory}.

Indeed, the theoretical calculation exhibits the existence of two, rather than one, specific heat
peaks.  We can now explain the origin of the peaks
as resulting from the
derivatives $\left(\frac{\partial p_c^s}{\partial P}\right)_T$ and $\left(\frac{\partial p_c^\ell}{\partial P}\right)_T$. These derivatives change most abruptly when the micro-clusters form (or dissolve), each
at a specific temperature determined by $(h_w^s-h_c^s)/\ln{g_w^s}$ and $(h_w^\ell-h_c^\ell)/\ln{g_w^\ell}$. Note that there can be pressures (both upper and lower boundaries) where the 
the sign of $(h_w^s-h_c^s)$ or $(h_w^\ell-h_c^\ell)$ change sign and the peak can be lost.

In summary, we have presented a simulational discovery of an unexpected second peak in the
temperature dependence of the specific heat of a popular model of glass-formation. This discovery means that the universal expectation of seeing a single ``specific heat peak" should be seriously revised. As there are two peaks in this examples, other examples may have multiple peaks or one peak.
Even in this system at a different pressure the two peaks may merge, giving the appearance of a single peak. We have also presented a detailed explanation of the existence of the two peaks, based on an approximate equation of state that we have derived on the basis of statistical mechanics. Indeed, it is not difficult to explain the two peaks in terms of the effects of micro-melting of clusters of small and than
of large particles when the temperature is increased. The important consequence is, however, that no universality is expected for the thermodynamics properties of different glass formers.

\acknowledgments

This work had been supported in part by the Israel Science Foundation, by the German Israeli Foundation and by the Minerva Foundation, Munich, Germany. 


\end{document}